\documentclass[prb,showpacs,twocolumn,amsmath,amssymb]{revtex4}

\usepackage{graphicx}
\usepackage{psfrag}
\usepackage{dcolumn}
\usepackage{bm}

\begin{document}

\newcommand{\ba}{\begin{array}}
\newcommand{\ea}{\end{array}}
\newcommand{\bc}{\begin{center}}
\newcommand{\ec}{\end{center}}
\newcommand{\nen}{\nonumber}
\newcommand{\eps}{\epsilon}
\newcommand{\dbar}{{\delta \!\!\!\! \smallsetminus}}
\newcommand{\tr}{\textrm{ tr}}
\newcommand{\sumint}{{\textstyle \sum}\hspace{-2.7ex}\int}
\newcommand{\atanh}{\textrm{atanh}}

\newcommand{\UA}{\uparrow}
\newcommand{\DA}{\downarrow}

\newcommand{\x}{\mathbf{x}}
\newcommand{\ist}{\!=\!}
\newcommand{\ket}[1]{{\left| #1 \right>}}
\newcommand{\bra}[1]{{\left< #1 \right|}}
\newcommand{\bracket}[2]{{\left< {#1} \left| {#2} \right.\right>}}
\newcommand{\bigbracket}[2]{{\big< {#1} \big| {#2} \big>}}
\newcommand{\Bigbracket}[2]{{\Big< {#1} \Big| {#2} \Big>}}
\newcommand{\bigket}[1]{{\big| #1 \big>}}
\newcommand{\bigbra}[1]{{\big< #1 \big|}}
\newcommand{\Bigket}[1]{{\Big| #1 \Big>}}
\newcommand{\Bigbra}[1]{{\Big< #1 \Big|}}
\newcommand{\me}[3]{{\left< {#1} \left| {#2} \right| {#3} \right>}}
\newcommand{\bigme}[3]{{\big< {#1} \big| {#2} \big| {#3} \big>}}
\newcommand{\bigmeS}[3]{{\big< {#1} \big| {#2} \big| {#3} \big>_S}}
\newcommand{\Bigme}[3]{{\Big< {#1} \Big| {#2} \Big| {#3} \Big>}}
\newcommand{\green}[1]{{ \left<\!\left< {#1} \right>\!\right> }}
\newcommand{\biggreen}[1]{{ \big<\!\big< {#1} \big>\!\big> }}
\newcommand{\Biggreen}[1]{{ \big<\!\big< {#1} \big>\!\big> }}

\newcommand{\define}{\stackrel{\textrm{\tiny def}}{=}}
\newcommand{\mustbe}{\stackrel{!}{=}}
\newcommand{\n}{\Hat{n}}
\newcommand{\cl}{{\cal L}}
\newcommand{\phd}{^{\phantom{\dagger}}}
\newcommand{\Dint}{\int {\cal D}[\SC]\;}
\newcommand{\YY}{{\cal Z}}

\newcommand{\cdag}{c^{\dagger}}
\newcommand{\cnod}{c^{\phantom{\dagger}}}
\newcommand{\fdag}{f^{\dagger}}
\newcommand{\fnod}{f^{\phantom{\dagger}}}
\newcommand{\bdag}{b^{\dagger}}
\newcommand{\bnod}{b^{\phantom{\dagger}}}

\newcommand{\Ueff}{{U_{\textrm{eff}}}}
\newcommand{\tUeff}{{\tilde U_{\textrm{eff}}}}
\renewcommand{\bottomfraction}{0.99}
\renewcommand{\textfraction}{0.01}


\title{Polaronic Quasiparticles in a Strongly Correlated Electron Band}

\author{W. Koller}\email{w.koller@imperial.ac.uk}
\author{A. C. Hewson}\email{a.hewson@imperial.ac.uk}
\author{D. M. Edwards}\email{d.edwards@imperial.ac.uk}
\affiliation{Department of Mathematics, Imperial College, London SW7 2AZ, UK
}

\date{18 July 2005}

\begin{abstract}
  We show that a strongly renormalized band of polaronic quasiparticle
  excitations is induced at the Fermi level of an interacting many-electron
  system on increasing the coupling of the electrons to local phonons. We give
  results for the local density of states at zero temperature both for the
  electrons and phonons. The polaronic quasiparticles satisfy Luttinger's
  theorem for all regimes considered, and their dispersion shows a kink
  similar to that observed experimentally in copper oxides. We calculate the
  quasiparticle weight factor $z$ and deduce the local effective
  inter-quasiparticle interaction $\tilde U$.
  Our calculations are based on the dynamical mean field theory and the
  numerical renormalization group for the hole-doped Holstein-Hubbard model
  and large on-site repulsion.
\end{abstract}
%

%
%

\pacs{71.10.Fd,71.30.+h,71.38.-k}

\keywords{polarons, Holstein model, Hubbard model, dynamical mean field
  theory, numerical renormalization group} 

\maketitle

\section{Introduction}

Polarons as quasiparticles, in which the electronic excitations are
strongly coupled to the lattice modes, have been studied  theoretically
for well over 50 years.
Nearly all these studies have been confined to models with only
one or two electrons, and do not include the electron spin. Such models
have limited applicability and there is a need for studies of more
realistic many-electron situations\cite{MR98a}. 
There are technical problems, however, in handling such
models as non-perturbative methods are required to describe the  effects of
strong electron-lattice interactions and few such methods are applicable to the
many-electron case. Progress has been made in recent years in dealing with
models with strong local interactions via the dynamical mean field theory
(DMFT)\cite{GKKR96,Jar92}.  
Calculations based on the Holstein and Holstein-Hubbard models, where
the electrons are coupled to local Einstein lattice modes, have been made
using the DMFT to examine the effect of strong electron-phonon interactions 
on the Mott transition\cite{KMH04,SCCG05}, for example.
The Holstein model\cite{Hol59}, however, was originally introduced to describe
small polaron effects. Within DMFT, only the single electron case can be
solved exactly\cite{Sum74,CPFF97}. There have been few studies at finite
electron densities and including the electron spin\cite{CC03}, except in one
dimension\cite{WF97,FWHWB04}. The problem is that if spin is included in the
Holstein model, there is an effective attractive local interaction between the
electrons, which leads to local bipolaron formation and, away from
half filling, superconductivity.

Once the electron spin is included the more realistic model is the
Holstein-Hubbard model which includes an local electron repulsion $U$. 
In most situations the repulsion $U$ dominates and undermines any tendency
to the formation of local bipolarons. In the Holstein-Hubbard model at
half filling with a large $U$, the electron-phonon interaction plays no
important role as charge fluctuations are suppressed; its main effect is to
renormalise and reduce the effective value of $U$\cite{KMH04,SCCG05}. At 
half filling and small $U$, where the electron-phonon coupling dominates,
again there is little evidence\cite{KMH04} of purely polaronic effects,
because of the formation of bipolarons\cite{MHB02,KMH04}. 
The question arises, therefore, whether there is any parameter regime where
purely polaronic effects can be observed. If there is, then does the behaviour
correspond to the predictions based on the spinless models with only one or
two electrons, or do many-electron correlations play a significant role?  

In this letter we present calculations to show clearly that small
strongly renormalized  polaronic excitations do exist in the Holstein-Hubbard
model and form a distinct narrow band at the Fermi level. 
The Hamiltonian for this model is    
\begin{equation}                                        \label{eq:hamil}
  \begin{aligned}
    H =&
    \sum_{{\bf k}\sigma}
    \epsilon({\bf k})\, \cdag_{{\bf k}\sigma} \cnod_{{\bf  k}\sigma} +
    U \sum_{i} n_{i\uparrow} n_{i\downarrow} \\ &+
    \omega_0 \sum_i  \bdag_i \bnod_i +
    g\sum_i  (\bdag_i + \bnod_i) \big(n_{i\UA}+ n_{i\DA}-1 \big) \:,
    \end{aligned}
\end{equation}
where  $\epsilon({\bf k})$ describes the dispersion of the band electrons and
$U$ is the on-site inter-electron repulsion as in the Hubbard model\cite{Hub63}.
The electron density $n_i = n_{i\UA} + n_{i\DA}$ at site $i$ couples linearly to
the local displacement operator $x_i \equiv (\bnod_i + \bdag_i)/\sqrt{2  m\omega_0}$ 
with an electron-phonon coupling $g$. 
The phonons are assumed to be dispersionless (local Einstein phonons) with
energy $\omega_0$ and $m$ is the mass of the vibrating ions.

Our calculations are based on DMFT using a semi-elliptical density of states
$\rho_0(\omega)$ of the non-interacting system given by
\begin{equation}                                      \label{eq:DOS}
  \rho_0(\omega) = \frac {2}{\pi D^2}\sqrt{D^2-(\omega+\mu_0)^2}
\end{equation}
for $|\omega+\mu_0| \leq D$ and zero elsewhere,
corresponding to a Bethe lattice, where $2D$ is the bandwidth and $\mu_0$ is
the chemical potential. The effective impurity problem is solved using the
numerical renormalization group (NRG)\cite{Wil75,KWW80a}, as generalized for
the calculation of dynamical response functions\cite{SSK92,CHZ94,BHP98}, and
incorporating the density matrix generalization\cite{Hof00} due to Hofstetter.

\begin{figure}
  \includegraphics[width=0.38\textwidth]{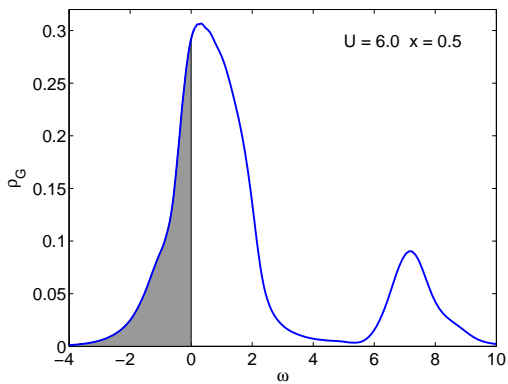}
  \caption{Local density of states for the quarter-filled Hubbard model ($U=6.0$) }
  \label{fig:g_spec_U60_x50}
\end{figure}

To suppress any tendency for bipolaron formation we take a large value of the
on-site repulsion, $U=6$ in units where $D=2$, which for half filling
corresponds to a Mott insulator~\cite{KMH04,Bul99}.
We choose a value of $\mu$  corresponding to quarter filling $\langle n_i\rangle=0.5$.
The local density of states for $g=0$ is shown in
Fig.~\ref{fig:g_spec_U60_x50}, with the Fermi level such that the lower
Hubbard band is $1/3$ filled.  In this regime the strong correlation
effects are largely suppressed due to the charge fluctuations;
there is no sharp quasiparticle peak at the Fermi level, which occurs for
the small doping regime near half filling, and there is a moderately reduced
quasiparticle renormalization factor $z \approx 0.65$. The small spectral weight
between the upper and lower bands is due to the tails in the broadening in the
higher energy peaks of the discrete NRG results.

\begin{figure}
  \includegraphics[width=0.45\textwidth]{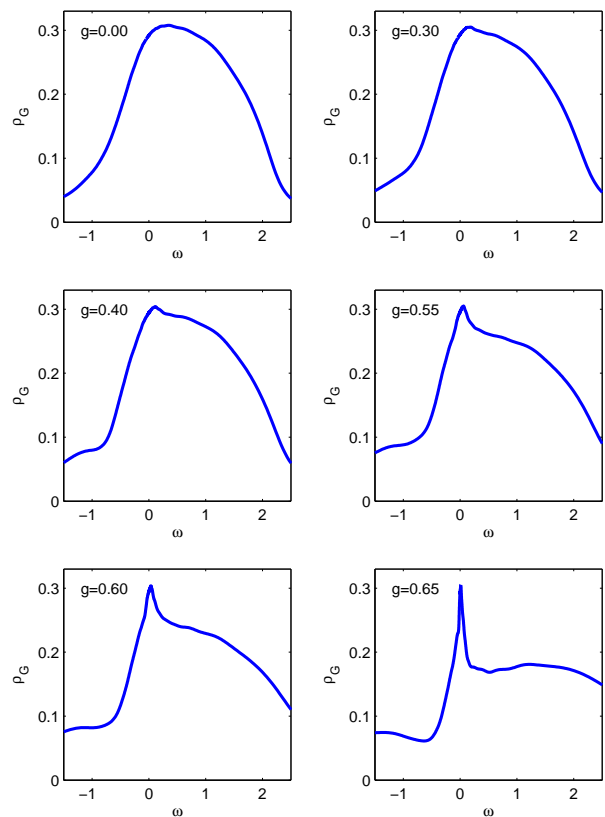}
  \caption{The development of the polaronic quasiparticle band in the lower
  Hubbard band of the quarter-filled Holstein-Hubbard model ($U=6.0$) for increasing
  electron-phonon coupling $g$.} 
  \label{fig:g_spec_U60_gx_x50}
\end{figure}

To study the impact of electrons coupling to phonons we concentrate on the
lower Hubbard band. 
In Fig.~\ref{fig:g_spec_U60_gx_x50} we show the effect of increasing the
electron-phonon interaction $g$, taking $\omega_0=0.2$ in all cases.
As $g$ increases to the value $g=0.65$ a sharp feature develops at the Fermi
level which we attribute to a polaronic renormalization. Due to the large $U$,
there is negligible double occupancy $\langle n_{i\uparrow}
n_{i\downarrow}\rangle < 0.01$ for this range of $g$ indicating no local
bipolaron formation.
\begin{figure}
  \includegraphics[width=0.40\textwidth]{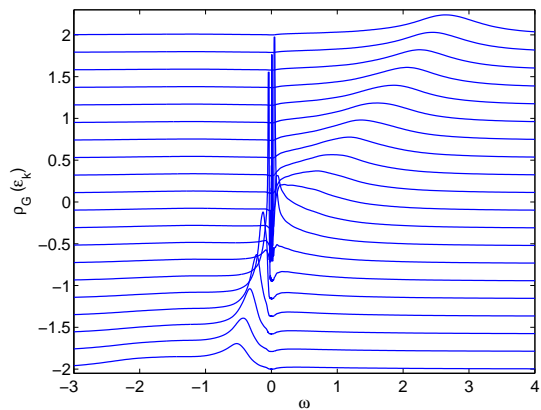}
  \caption{Spectral density $\rho_k(\omega)$ of the electron Green's function
  for the quarter-filled Holstein-Hubbard model ($U=6.0, g=0.60$). } 
  \label{fig:g_spec_k_U60_g60_x50}
\end{figure}

\begin{figure}
  \includegraphics[width=0.35\textwidth]{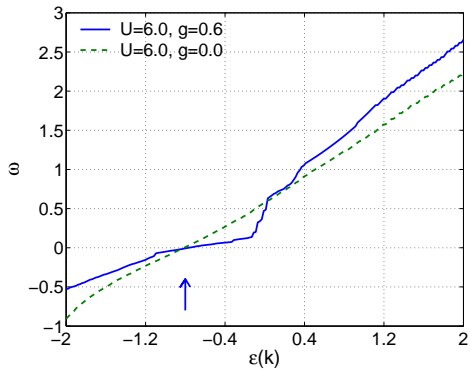}
  \caption{Positions of the maxima of the spectral function $\rho_k(\omega)$
  of the electron Green's function for the quarter-filled Holstein-Hubbard
  model ($U=6.0, g=0.60$, full line) and the quarter-filled Hubbard model
  ($U=6.0, g=0.00$, dashed line).
  The arrow indicates $\epsilon({\bf k}_F)$.}
  \label{fig:qp_band_U60_gx_x50}
\end{figure}

To look at the band aspects of these excitations, we introduce
the ${\bf k}$-resolved one-electron Green's function
$G_{{\bf k},\sigma}(\omega)$ with a local self-energy
$\Sigma_{\sigma}(\omega)$, 
\begin{equation}                                        \label{eq:gf}
  G_{{\bf k},\sigma}(\omega)=
  \frac {1}{\omega+\mu-\epsilon({\bf k})-\Sigma_{\sigma}(\omega)}\:.
\end{equation}
Figure~\ref{fig:g_spec_k_U60_g60_x50} shows a plot of the spectral density of
the Green's function $G_{\bf k}(\omega)$ for $-2 \leq \epsilon({\bf k}) \leq
2$ and $g=0.60$. When integrated over ${\bf k}$, it gives the local density of
states for the corresponding $g$ shown in Fig.~\ref{fig:g_spec_U60_gx_x50}.
We plot the positions of the maxima in the quasiparticle peaks as a function
of $\epsilon({\bf k})$ in Fig.~\ref{fig:qp_band_U60_gx_x50}. The flat
dispersion in the region of the Fermi level is clear evidence of a narrow 
polaronic band (full line). We see a kink at $\omega \approx -0.05$ which we
discuss later in relation with the phonon spectra. None of this is seen in the
pure Hubbard case, shown as the dashed line in Fig.~\ref{fig:qp_band_U60_gx_x50}.

The locus of points ${\bf k}_{\rm F}$ on the Fermi surface of the interacting
system is given by
\begin{equation}                          \label{eq:FS}
  \epsilon({\bf k}_{\rm F})=\mu-\Sigma(0) \: .
\end{equation}  
According to Luttinger's theorem\cite{Lut60} the volume of the Fermi surface
should be the same as that of the non-interacting system. For a local self
energy~$\Sigma(\omega)$, this implies that the Fermi surface given by
Eq.~(\ref{eq:FS}) should be the same as the non-interacting one,
$\epsilon({\bf k}_{\rm F})=\mu_0$. If these are to coincide, then
$\mu_0 = \mu - \Sigma(0)$.

\begin{figure}
  \includegraphics[width=0.45\textwidth]{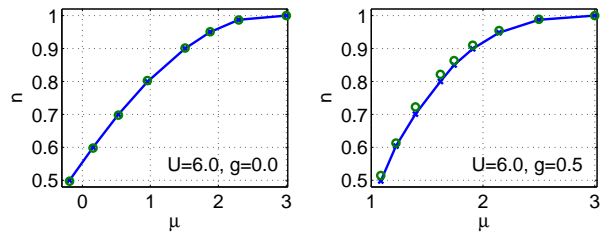}
  \caption{Density  as a function of the chemical potential $\mu$
    calculated directly (full lines) and from the volume of the Fermi surface
    (circles) for the Hubbard model (left-hand plot, $U=6$) and the
    Holstein-Hubbard model (right-hand plot, $U=6, g=0.5$).}
  \label{fig:n_vs_mu_gx.eps}                 
\end{figure}

To check the theorem we have calculated the density $\langle n_i\rangle$ 
directly from the NRG expectation value in the ground state, and then from
the non-interacting Fermi surface using the value of $\mu_0 = \mu - \Sigma(0)$
from equation (\ref{eq:FS}). The two sets of results are compared for a range
of hole dopings in Fig.~\ref{fig:n_vs_mu_gx.eps}~(left) for the pure Hubbard
model, and in Fig.~\ref{fig:n_vs_mu_gx.eps}~(right) for the Holstein-Hubbard
model with $g=0.5$. The agreement between the two sets of results provide a
convincing evidence of the applicability of Luttinger's theorem in strongly
correlated systems both with and without the electron-phonon interaction. 

If we assume Fermi liquid theory\cite{Lut60} and expand the self-energy in
Eq.~(\ref{eq:gf}) about the Fermi level $\omega=0$, and retain the first two
terms, we can define a quasiparticle Green's function 
\begin{equation}                          \label{eq:qp_gf}
  \tilde G_{{\bf k},\sigma}(\omega)=\frac{1}{\omega+\tilde\mu-\tilde\epsilon({\bf k})}
\end{equation}
where $\tilde\epsilon({\bf k})=z\epsilon({\bf k})$, and $z$ is the usual
wavefunction renormalization factor $z=(1-\Sigma'(0))^{-1}$, and  $\tilde
\mu=z(\mu-\Sigma(0))$ is a renormalized chemical potential. 
The corresponding density of states $\tilde \rho_0(\omega)$ for the
non-interacting quasiparticles is given by 
\begin{equation}                          \label{eq:qp_DOS}
  \tilde\rho_0(\omega) = \frac 2 {\pi {\tilde D}^2}
  \sqrt{{\tilde D}^2-(\omega+\tilde \mu)^2}
\end{equation}
where $\tilde D=zD$ plays the role of a renormalized band width. We interpret
this as a renormalized band of free polaronic quasiparticles. Integrating
this free quasiparticle band up to the Fermi level gives the
quasiparticle occupation $\langle \tilde n_i\rangle$.
The resulting $\langle \tilde n_i\rangle = \langle n_i\rangle$, demonstrating
the one-to-one correspondence of the electron and quasiparticle states of
Fermi liquid theory\cite{Lan57a,Lan57b} (see  Fig.~\ref{fig:n_vs_mu_gx.eps}).

\begin{figure}
  \includegraphics[width=0.35\textwidth]{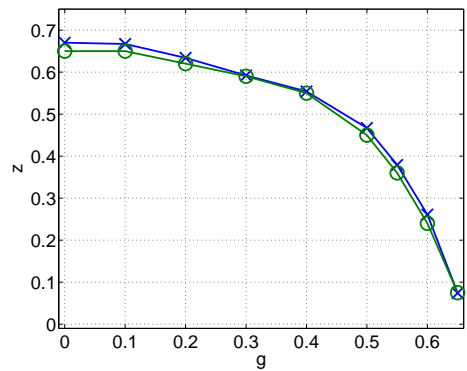}
  \caption{Quasiparticle weight as a function of $g$ for the quarter-filled
    Holstein-Hubbard model ($U=6.0$) deduced from the derivative of the self
    energy ($\circ$) and from an analysis of the renormalized parameters
    ($\times$).} 
  \label{fig:z_U60_gx_x50}
\end{figure}

In Fig.~\ref{fig:z_U60_gx_x50} we give the results for the change in the
quasiparticle renormalization factor $z$ with increase of $g$ for quarter
filling and $U=6$. The values of $z$ we calculated by differentiating the
self-energy, and also from an analysis of the renormalized
parameters\cite{HOM04,Hew05} on the approach to the low energy fixed point.
The two types of estimates are in reasonable agreement; the deviation for
smaller values of $g$ is largely due to fluctuations in the
numerical differentiation of $\Sigma(\omega)$ in this range. The behaviour of
$z$ with increase of $g$ is quite different from that near half filling where
$z$ increases with increase of $g$. 
Near half filling the strong correlation effects stem from the enhancement of
spin fluctuations and the suppression of charge fluctuations due to the
large value of $U$. In this regime the increase of $z$ with $g$ can be
understood as due to the reduction in the effective value of $U$, $U_{\rm
eff}=U-2g^2/\omega_0$;  hence an increase in $g$ reduces  $U_{\rm
eff}$ and hence reduces the renormalization\cite{KMH04,SCCG05}. The results
for $z$ at quarter filling shown in Fig.~\ref{fig:z_U60_gx_x50}, by contrast,
show a large decrease of $z$ with increase of $g$. 
This increased renormalization is a pure polaronic effect;
the coupling to the lattice causing an increase in the effective mass of the
conduction electrons and a reduced mobility.

\begin{figure}
  \psfrag{U tilde}{$\:\:\widetilde U$}
  \psfrag{g}{$\:\:g$}
  \includegraphics[width=0.32\textwidth]{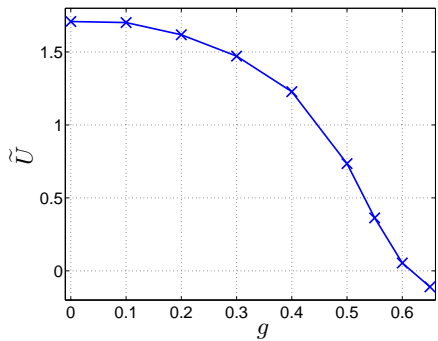}
  \caption{Effective local interaction $\tilde U$ between quasiparticles as
    a function of $g$ for the quarter-filled Holstein-Hubbard model
    ($U=6.0$).} 
  \label{fig:Utilde_U60_gx_x50}
\end{figure}

The renormalized parameter approach\cite{HOM04,Hew05} also allows us to
calculate the local interaction between the quasiparticles $\tilde U$. The
results for $\tilde U$ over this range are shown in
Fig.~\ref{fig:Utilde_U60_gx_x50}. There is a progressive reduction with
increase of $g$ as might be expected for the induced attactive interaction due
to phonon exchange. 
For the largest value of $g$ the value of $\tilde U$ becomes negative,
even though $U-2g^2/\omega_0>0$.
This can be interpreted as due to the fact the very low energy quasiparticles
hardly ever occupy the same site due to correlation effects (double occupancy
$<0.01$) and their effective attraction is due to retardation effects.

\begin{figure}
  \includegraphics[width=0.35\textwidth]{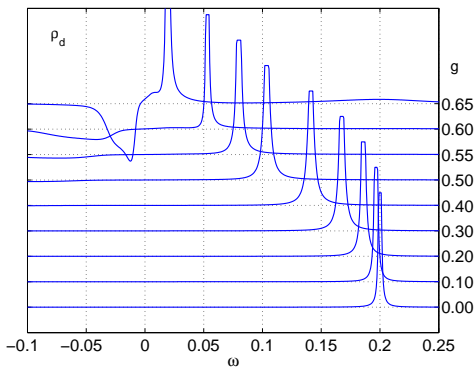}
  \caption{Spectral density of the phonon propagator for the quarter-filled
  Holstein-Hubbard model ($U=6.0$) for various values of the electron-phonon
  coupling.} 
  \label{fig:b_spec_U60_gx_x50}
\end{figure}

For values of $g>0.65$ the dynamical mean field theory no longer
converges satisfactorily, the iterations fluctuate between occupied and
unoccupied states. This indicates the on-set of a broken symmetry
state, probably some form of charge ordering or phase separation. This
instability is also revealed in the one-phonon propagator $d(\omega) =
\biggreen{\bnod_i\,;\bdag_i}_\omega$ whose spectral density is shown in 
Fig.~\ref{fig:b_spec_U60_gx_x50}.
With increasing $g$ there is a progressive softening of the phonon mode from
its bare value $\omega_0=0.2$. For $g\approx 0.65$ considerable spectral
weight develops for $\omega < 0$. This weight relates to the expected number
of excited phonons in the ground state. It clearly indicates an incipient
lattice instability for $g>0.65$. 

The kink to be seen at $\omega=-0.05$ in the quasiparticle dispersion in
Fig.~\ref{fig:qp_band_U60_gx_x50} is clearly related to the frequency of the
renormalized phonon excitation at the corresponding $g=0.6$
(Fig.~\ref{fig:b_spec_U60_gx_x50}). 
An analysis for different values of $g$ shows a clear correlation of the kink
energy to the {\em renormalized} phonon frequency. This is similar to the
experimentally observed kink in the hole-doped copper oxides\cite{Lea01}.

To summarize: We have shown that an accurate treatment of the Holstein-Hubbard
model at {\em finite electron density} and $T=0$ reveals a clear polaronic
quasiparticle band induced at the Fermi level. Luttinger's theorem is shown to
be satisfied. A kink found in the quasiparticle dispersion relation is related
to the renormalized phonon frequency.


\begin{acknowledgments}

We wish to thank the EPSRC (Grant GR/S18571/01) for financial
support.

\end{acknowledgments}

\end{document}